\documentclass[twocolumn,nofootinbib,amsmath,amssymb,letter]{revtex4}

\usepackage{graphicx}
\usepackage{dcolumn}
\usepackage{bm}

\DeclareMathOperator {\re}{Re}
\DeclareMathOperator {\im}{Im}

\begin{document}

\title{Angles from B Decays with Charm: Summary of Working                               
Group 5 of the CKM Workshop 2006 }

\author{Gianluca Cavoto}
 \email{gianluca.cavoto@roma1.infn.it}
\affiliation{%
Universit\`a di Roma La Sapienza,  INFN and Dipartimento di Fisica, I-00185 Roma, Italy \\
}%

\author{Robert Fleischer}
 \email{robert.fleischer@cern.ch}
\affiliation{%
Theory Division, Department of Physics, CERN, CH-1211 Geneva 23, Switzerland\\
}%

\author{Karim Trabelsi}
 \email{karim.trabelsi@kek.jp}
\affiliation{%
Institute of Particle and Nuclear Studies, KEK,\\
1-1 Oho, Tsukuba-shi, Ibaraki-ken, 305-0801, Japan\\
}%

\author{Jure Zupan}
 \email{jure.zupan@ijs.si}
\affiliation{Department of Physics, University of Ljubljana, Jadranska 19, 1000
Ljubljana, Slovenia}
\affiliation{J. Stefan Institute, Jamova 39,
P.O. Box 3000, 1001 Ljubljana, Slovenia}

\begin{abstract}
 We summarize the results presented in Working Group 5 (WG5) of the CKM 
 2006 Workshop in Nagoya. The charge of WG5 was to discuss the measurements 
 of unitarity triangle angles $\beta/\phi_1$ and $\gamma/\phi_3$ from $B$-meson 
 decays containing charm quark(s) in the final states. 
 \end{abstract}

\maketitle

\section{Introduction}

The focus of Working Group 5 at the CKM 2006 Workshop were the measurements of 
$\beta/\phi_1$ and $\gamma/\phi_3$ 
angles in the standard unitarity triangle 
of the Cabibbo--Kobayashi--Maskawa (CKM) matrix \cite{bib:C,bib:KM} 
that are obtained from $B$ decays into final state mesons with valence 
$c$ quark(s). The discussion summary of the previous edition in this 
workshop series can be found in \cite{CKM05}.

\section{Measurements of $\beta$ in B decays with charmonium.}\label{sec:beta}
$B^0$-meson decays originating from $b \to c \bar c s$ quark-level transitions 
are the key channels to measure the $B^0_d$--$\bar B^0_d$ mixing phase
\cite{bisa}. In the CKM picture of CP violation this phase equals $2\beta/\phi_1$, with
\begin{equation}
\beta = \mbox{Arg}\left(-\frac{V_{cd}V_{cb}^*}{V_{td}V_{tb}^*}\right), 
\end{equation}
one of the angles in the standard unitarity triangle. A particularly clean measurement
of $\beta$ is provided by the time dependent CP asymmetry in the ``golden" channel $B^0 \to J/\psi K_S$.
Here the theory error, i.e. the difference between $\sin(2\beta)$ and the coefficient of $\sin(\Delta m t)$, $S_{J/\Psi K_S}$,
is small because it is given by a doubly Cabibbo-suppressed ratio of penguin to tree amplitudes \cite{RF-rev}. Neglecting
this correction leads to small ``penguin pollution" of extracted $\sin(2\beta)$.

In view of the steadily increasing accuracy at the $e^+e^-$
$B$ factories and the quickly approaching start of the LHC, a closer look at the size of penguin pollution
in $B^0 \to J/\psi K_S$ was taken as part of our Working Group discussions. The conclusion of an analysis performed
several years ago \cite{BMR}  was that these corrections
are extremely small, of the order of less than a per mil of the observed value,
although a precise calculation is not possible. The analysis was extended recently by the authors of \cite{LM} using a formalism that combines the QCD-improved factorization and the perturbative QCD approaches. The penguin pollution $\Delta S_{J/\psi K_S}$ and the direct CP asymmetry $A_{J/\psi K_S}$ were calculated
at leading power in $1/m_b$ and at next-to-leading order in $\alpha_s$.
Both quantities were found to be at the $10^{-3}$ level \cite{ref:mishima}. 

A different avenue to deal 
with these corrections was chosen by the authors of \cite{ref:ciuchini,CPS}: employing the $SU(3)$
flavour symmetry of strong interactions and further plausible dynamical assumptions, the data from $B^0\to J/\psi \pi^0$ channel, where penguin-to-tree ratio is not
CKM suppressed, are used to estimate $\Delta S_{J/\psi K_S}$. 
A fit to the current data gives $\Delta S_{J/\psi K_S}=
0.000 \pm 0.012$. This estimate of $\Delta S_{J/\psi K_S}$
is an order of magnitude larger than the alternative ones discussed above, and is
comparable to the present experimental systematic error. Note, however, that the quoted error reflects also the size of experimental errors on observables in $B^0\to J/\psi \pi^0$ decay and does not necessarily reflect the size of penguin pollution. In this sense the quoted bound on penguin pollution is a conservative one.
At the LHC, the penguin
pollution in $B^0_d \to J/\psi K_S$ can be controlled using the $B^0_s \to J/\psi K_S$
channel and the $U$-spin symmetry \cite{RF-BspsiK} as sketched below. 

The above estimates of penguin pollution are especially interesting in light of a rather small recent experimental $\sin(2\beta)$ value, which leads to some tension in the CKM fits. Following a recent paper \cite{BF-06} the implications for the allowed region in the 
space of the general New Physics (NP) parameters for $B^0_d$--$\bar B^0_d$ mixing were discussed  \cite{ball}.
In this analysis, the ``true" value of $\beta$ is fixed by 
$\gamma$ and $|V_{ub}|$ extracted from tree-level processes, which are 
assumed not to be affected by NP. Comparison with the value of
$\beta$ extracted from 
$B^0_d \to J/\psi K_S$ then gives a constraint on a NP phase $\phi_d^{\rm NP}$
in $B^0_d$--$\bar B^0_d$ mixing. The result depends sensitively 
on $|V_{ub}|$, where the inclusive and exclusive determinations give 
$\phi_d^{\rm NP}|_{\rm incl}=-(11.0\pm4.3)^\circ$ and
$\phi_d^{\rm NP}|_{\rm excl}=-(3.4\pm7.9)^\circ$, respectively. Similar effects 
were also found in Refs.~\cite{NP-Bd}, and should be closely monitored in the future.

%
 
\begin{figure}[hbtp]
\begin{center}
\includegraphics[width=0.5\textwidth]{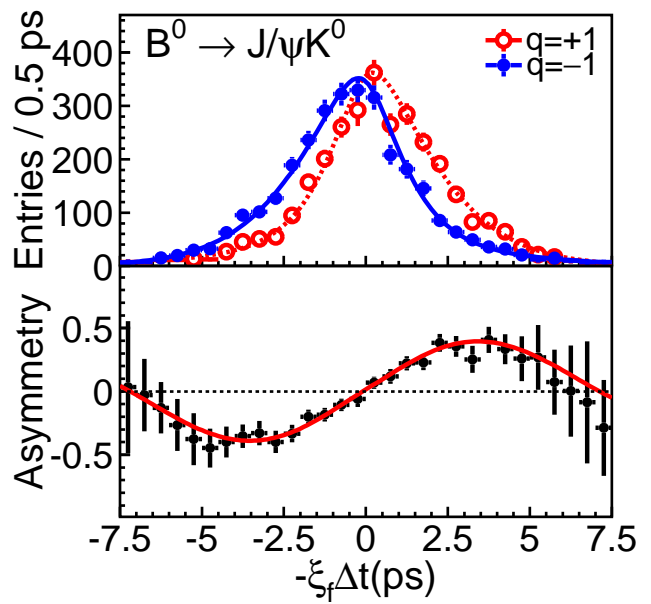}
\end{center}
\caption{
Background subtracted $\Delta t$ distributions and asymmetry for events with good tags 
for $J/\psi  K^0$ modes in the Belle analysis. In the asymmetry plot, solid curve shows the 
fit result.}
\label{fig:sin2betabelle}
\end{figure}

\begin{figure}[hbtp]
\begin{center}
\includegraphics[width=0.4\textwidth]{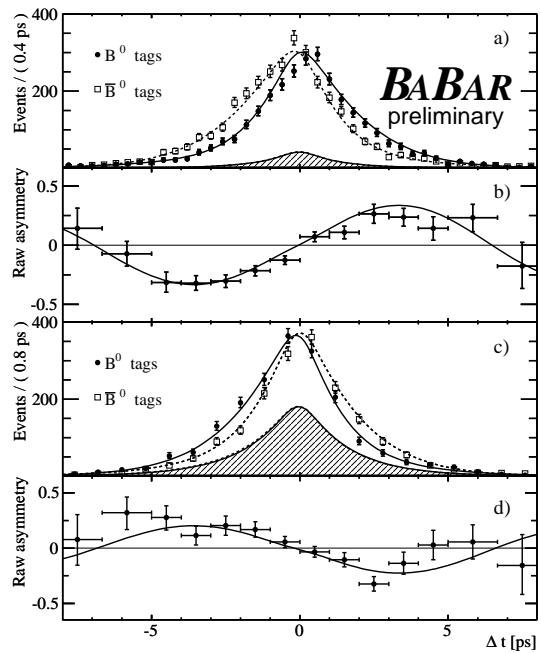}
\end{center}
\caption{
a) Number of  CP-odd  candidates ($J/\psi  K_S$, $\psi(2S) K_S$, $ \chi_{c1} K_S$, and $\eta_c K_S$)
in the signal region with a  $B^0$  tag ($N_{B^0 }$) and with a  $\bar B^0$  tag ($N_{ bar B^0}$), and 
b) the raw asymmetry $(N_{B^0}-N_{\bar B^0})/(N_{B^0}+N_{\bar B^0})$, as functions of $\Delta t$.
Figures c) and d) are the corresponding distributions for the CP-even  mode $J/\psi K_L$.  
 The solid (dashed) curves 
represent the fit projections in $\Delta t$ for $B^0$  ($\bar B^0$) tags. The shaded regions represent the estim
ated background contributions.}
\label{fig:sin2betababar}
\end{figure}

  On the experimental side, the $B$-factory experiments BaBar and Belle have analyzed datasets of 
384 and 535 $\times 10^6$ $B \bar B$ pairs, respectively.  The
  preliminary BaBar result is given as an average over several $c \bar c $ $K_S$/$K_L$  channels \cite{kathy}:  $\sin{2\beta} =  0.710 \pm 0.034 \pm 0.019$ and $ |\lambda| = 0.932 \pm 0.026 \pm 0.017$ (Fig.~\ref{fig:sin2betababar}). Belle reported the result using only the $J/\psi $ $K^0$ modes~\cite{bellesin2beta}: $\sin{2\phi_1} =  0.642 \pm 0.031 \pm 0.017$ and $ A = 0.018 \pm 0.021 \pm 0.014$ with $A = - C$  $=$ $({1 - |\lambda|^2})/({1 + |\lambda|^2})$ (Fig.~\ref{fig:sin2betabelle}).  It should  be noticed that final states with different ($c \bar c $) resonances can have different  $\Delta S$ corrections. At present these are expected to be smaller than experimental errors. With increasing experimental accuracy, however, averaging the CP asymmetry measurements from different modes may become problematic.   
Experimentally, the predicted uncertainty on  $S$ and $C$ from the B-factories at  $2$ ab${}^{-1}$
   will still be dominated by statistics while the systematic component of  the error originates mainly from the knowledge of the vertexing algorithm performance.

\subsection{Measurements of $\beta$ and cos2$\beta$.}

 Several methods based on the decays into resonant or multi-body final states are being used to resolve the discrete ambiguity in the determination of $\beta$ from measured $\sin 2\beta$. 
  A  theoretical review of $\beta$ determinations from $ B $ decays involving charm final states was presented   by A. Datta \cite{Datta:2007mj}: the $b\to c\bar c s$ transitions $B \to J/\psi K^{(*)} $, $B \to D^{(*)}\bar{D}^{(*)} K_S$, the $b \to c\bar{c} d$ transition  $B \to D^{(*)}\bar{D}^{(*)}$ and the $b \to c \bar{u} d$ transition $B \to D^{(*)}h^0$.

The decay $B \to J/\psi K^{0*} \to J/\psi K_S\pi^0$ is a VV decay. The corresponding 
time-dependent angular distribution allows a measurement of both $\sin 2\beta$ and 
$\cos 2\beta$ \cite{DDF-ambig}. The sign ambiguity can be resolved  by using the interference between 
the $K\pi$ $S$-wave and $P$-wave amplitudes in the $K^*(892)$ region and assuming small strong interactions between $J/\psi$ and $ K \pi$. The result of such an analysis yields a positive value for $\cos2\beta$ \cite{prd1}.

The $B(t) \to D^{*+}D^{*-} K_{\rm S}$ decay can have both non-resonant and resonant contributions making it sensitive to $\cos2\beta$ \cite{browder}. Using the theoretical calculation for the sign of the hadronic coefficient in front of $\cos2\beta$ 
\cite{browder}, $\cos2\beta$ is preferred to be positive at the 94\,\% confidence 
level \cite{babarDDK}.

The tree level $b\to c\bar{u}d$ decay $B(t) \to D h^0 (h^0=\pi^0, \eta ...)$ with $D \to K_S \pi^+ \pi^-$ uses the variation of the strong phase over the final phase space to obtain $\beta$ without discrete ambiguities \cite{bondar}. The sensitivity to the phase comes from the interference of different resonance decays that either come from $B^0$ directly or from a prior oscillation through $\bar B^0$. Results of such analyses are available from both BaBar and Belle ~\cite{kathy,gaz}.

In order to obtain $\beta$ from $\bar{B}(t) \to D^{(*)}\bar{D}^{(*)} $, 
one needs further information to deal with the penguin effects. This can be
provided by the $U$-spin-related $ \bar{B_s} \to {D}_s^{(*)}\bar{D}_s^{(*)}$ 
modes \cite{RF-BspsiK,RF-BDD} or by using $SU(3)$-related 
$ \bar{B} \to {D}^{(*)}\bar{D}_s^{(*)}$ decays and dynamical assumptions \cite{justin}.
The BaBar measurements of CP observables in $\bar{B}(t) \to D^{(*)+}\bar{D}^{(*)-} $ were presented  \cite{gaz}, while a new Belle measurement of $B(t)\to D^+D^-$ was reported \cite{fratina}. There is a slight  disagreement between the two experiments on the size of the direct CP asymmetry in  $B(t)\to D^+D^-$. While Belle obtains $C_{D^+D^-}  =-0.91 \, \pm 0.23 \,\pm 0.06$, BaBar quotes $C_{D^+D^-} =  0.11 \pm 0.35 \pm 0.06$. In the Standard Model (SM), a small direct CP asymmetry is expected based on an estimate using a combination of naive factorization and an arbitrarily large strong phase due to the final-state interactions \cite{Chen:2005rp}. In Ref.~\cite{RF-BDD}, a detailed analysis of the allowed region in observable space for CP violation in 
$B^0_d \to D^+ D^-$ was performed in view of the new $B$-factory measurements,
together with an estimate of the relevant hadronic penguin parameters and observables. The questions of the most promising strategies for the extraction 
of CP-violating phases, about the interplay with other measurements of CP violation 
and regarding NP search were also addressed. 


\subsection{CPT/T violation in mixing.}


 BaBar  has presented experimental results on $CP$ and $CPT$ violation in mixing \cite{Covarelli:2007cn}. Allowing for $CPT$ violation, the general parametrization of 
 $B^0$--$\bar B^0$ mixing is
\begin{eqnarray}
|B_L\rangle&=&p\sqrt{1-z}|B^0\rangle +q\sqrt{1+z}|\bar B^0\rangle\nonumber ~ ,\\
|B_H\rangle&=&p\sqrt{1+z}|B^0\rangle -q\sqrt{1-z}|\bar B^0\rangle ~,
\label{eq:mass_eigenstates_cpt}
\end{eqnarray}
with $z$ denoting a complex parameter that is zero if $CPT$ is conserved. On the other hand, $CP$ violation in mixing is found, if $|q/p|\ne 1$ (while $CP$ violation in the interference of mixing and decay is possible if $\arg(q/p)\ne 0$).

In the SM, $|q/p|$ is close to 1,
\begin{equation}
\left|\frac qp\right|-1 \approx -\frac{1}{2}\im \frac{\Gamma_{12}}{M_{12}},
\end{equation}
where $\langle B^0|H_{\rm eff}|\bar B^0\rangle=M_{12}-i \Gamma_{12}/2$.
The $CP$-violating quantity $\im (\Gamma_{12}/M_{12})$ is suppressed by an additional factor $(m_c^2-m_u^2)/m_b^2\approx 0.1$ relative to $|\Gamma_{12}/M_{12}|$, giving \mbox{$|\im ({\Gamma_{12}}/{M_{12}})|<10^{-3}$} in the SM~\cite{ciuch,bene}.

The $CP$- and $CPT$-violating parameters are determined from time-dependent fits to $B^0$--$\bar B^0$ pair events in two complementary approaches. 
 In the first approach, two high-momentum leptons are demanded in order to
 select inclusive semileptonic $B^0$ decays.
In the second approach, one of the $B$ mesons is partially reconstructed in the semileptonic $D^{*-}l^+\nu_l$ channel  (only the lepton and the soft pion from $D^{*0}\to \bar D^0 \pi^-$ decay are reconstructed), while for the flavour of the other 
$B$ a leptonic tag is used.

\begin{figure}
\begin{center}
\includegraphics[width=0.4\textwidth]{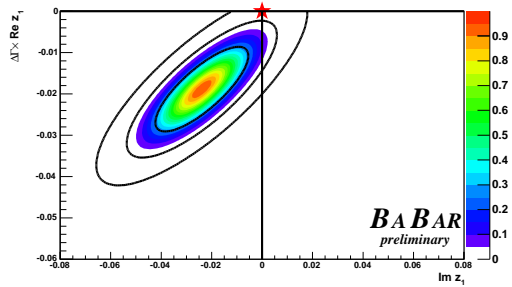}
\end{center}
\caption{Allowed regions for the $CPT$-violating   parameters $\im z_{1}$ and $\Delta \Gamma \re z_1$ at various confidence levels. The red star represents the SM expectation and the solid black ellipses correspond to 1, 2 and 3$\sigma$ significances.}
\label{fig:CPTv}
\end{figure}

No evidence of $CP$ or $CPT$ violation is found in mixing with either of 
the two methods. The first method gives
\begin{eqnarray*}
|q/p| -1 & = & (-0.8 \pm 2.7_{\rm (stat.)}  \pm 1.9_{\rm(syst.)})\times  10^{-3},\\
\im z  & = & (-13.9  \pm 7.3_{\rm (stat.)} \pm 3.2_{\rm(syst.)})\times  10^{-3},\\
\Delta \Gamma \re z & = & (-7.1 \pm 3.9_{\rm (stat.)} \pm 2.0_{\rm(syst.)})\times 10^{-3}{\rm ps}^{-1},
\end{eqnarray*}
where $z$ was taken to be time independent. 
The preliminary result from the second method is:
\begin{equation}
|q/p| -1 = (6.5 \pm 3.4_{(\mathrm{stat.})} \pm 2.0_{(\mathrm{syst.})}) 
\times 10^{-3}. 
\end{equation}
Both results are compatible with the SM expectations and with previously published BaBar results~\cite{fmv,bozzi}. As first pointed out by Kosteleck\`{y} \cite{koste}, taking the $CPT$-violating parameter $z$ to be constant in time is not very natural. Since 
$CPT$ violation in the quantum field theory implies Lorentz violation  one can expect 
$z \propto \beta^{\mu}\Delta a_{\mu}$, where $ \beta^{\mu}$ is the decaying $B$-meson four-velocity and $\Delta a_{\mu}$ a constant four-vector describing Lorentz violation. Because of the Earth's rotation the product of the two vectors is time dependent
$ z = z_0 + z_1\cos{(\Omega\hat{t} + \phi)} $,
with $\Omega $ the Earth's rotation frequency, $ \hat{t}$ the sidereal time, while $z_0$ and $z_1$ are constants. BaBar analysis accounting for this time dependence gives results for $\im z_{1}$ and $\Delta \Gamma \re  z_1$ consistent with zero at 
$2.2 \sigma $ as shown in Fig. \ref{fig:CPTv}.

%



%

%

%



\section{Measurements of $\gamma$}

The extraction of $\gamma$ from $B\to (f)_D K$ decays uses the interference between $\overline{b} \to \overline{c} u \overline{s}$ and 
$\overline{b} \to \overline{u} c \overline{s}$ transitions. The interference is nonzero when the final state $f$ is accessible to both $D$ and $\overline{D}$  mesons. The theoretical uncertainty is completely negligible as there are
no penguin contributions.

Several methods 
were proposed that differ in the choices for the final states $f$: 
$C\!P$ eigenstate (GLW method~\cite{GLW}),
doubly Cabbibo suppressed (ADS method~\cite{ADS}), and a combination of these two methods
using a $D$ Dalitz analysis (GGSZ method ~\cite{GGSZ}). 

The feasibility of the $\gamma$ measurement crucially depends 
on the size of $r_B$, the ratio of the $B$ decay amplitudes involved 
($r_B =  {|A(B^+ \to \overline{D} K^+)|}/{ |A(B^+ \to  D K^+)|}$). 
The value of $r_B$ is given by the ratio of the CKM matrix elements 
$|V_{ub}^{*} V_{cs}^{}|/|V_{cb}^{*} V_{us}^{}|$
and the colour suppression factor, and is estimated to be in the range 
0.1--0.2~\cite{GRONAU}. 
For different $D$ decays, the $B$ system parameters are common, which 
means that the  combination of  different $D$ channels can help more  than just adding more
statistics \cite{Gronau:2004gt}. 

The $\Delta \gamma$ shift due to $D$--$\overline{D}$ mixing is estimated to be less than
one degree for doubly Cabibbo-suppressed decays and much smaller in 
other cases, and can eventually be included in the $\gamma$ determination.
The effect due to $C\!P$ violation in the neutral $D$ sector 
is negligible in the SM and at most at the $10^{-2}$ order
if one considers NP in the charm sector \cite{Grossman:2005rp}.



Results from the two $B$-factories Belle/KEKB and BaBar/PEPII are available.
The Belle collaboration uses a data sample that consists 
of $386 \times 10^6 B\bar{B}$ pairs~\cite{DALITZ_BELLE}. 
The decay chains $B^+ \to D K^+$,
$B^+ \to D^* K^+$ with $D^* \to D\pi^0$ and $B^+ \to D K^{*+}$ with 
$K^{*+} \to K^0_S \pi^+$ are selected for the analysis.  The analysis of the
BaBar collaboration~\cite{DALITZ_BABAR_NEW} is based on 
$347 \times 10^6 B\bar{B}$ 
pairs. The reconstructed final states are $B^+ \to D K^+$ and 
$B^+ \to D^* K^+$ with two $D^*$ channels: $D^* \to D\pi^0$ and 
$D^* \to D\gamma$.~\footnote{The previous BaBar~\cite{DALITZ_BABAR_OLD} publication 
includes also the $B^+ \to D K^{*+}$ channel but this mode is not included 
in the recent update.} The neutral
$D$ meson is reconstructed in the $K^0_S \pi^+ \pi^-$ final state in all
cases. 
The number of reconstructed signal events in the Belle's data are 
$331 \pm 23$, $81 \pm 11$ and $54 \pm 8$ for the $B^+ \to D K^+$,
$B^+ \to D^* K^+$ and $B^+ \to D K^{*+}$ channels, respectively. BaBar
finds $398 \pm 23$, $97 \pm 13$ and $93 \pm 12$ signal events in the 
$B^+ \to D K^+$, $B^+ \to D^*[D\pi^0] K^+$ and $B^+ \to D^*[D\gamma] K^+$ 
channels respectively.

The $\bar{D}^0 \to K^0_S \pi^- \pi^+$ decay amplitude $f(m^2_+, m^2_-)$ ($m^2_{\pm} = m^2(K^0_S \pi^{\pm}$) is determined independently from a large sample of flavor-tagged
$D^{*-} \to \bar{D}^0 \pi^-$, $\bar{D}^0 \to K^0_S \pi^+ \pi^-$ decays
produced in continuum $e^+ e^- \rightarrow q \bar q$ annihilation. 
The amplitude $f$ is parametrized as a coherent sum of two-body
decay amplitudes plus a non-resonant decay amplitude,
\begin{equation} \label{ffunc}
f(m^2_+, m^2_-) = \sum_{j=1}^{N} a_j e^{i\alpha_j} {\cal A}_j(m^2_+, m^2_-) +
b e^{i \beta},
\end{equation}
where the sum is over the resonances present in $K^0_S \pi^+ \pi^-$, ${\cal A}_j(m^2_+, m^2_-)$ is the
corresponding Breit-Wigner form, $a_j$ and $\alpha_j$ are respectively the amplitude and phase of the
matrix element for a decay through $j$-th resonance, 
while $b$ and $\beta$ are the amplitude and phase of the non-resonant component. 
The total phase and amplitude are arbitrary. To be consistent with the CLEO 
analysis~\cite{DALITZ_CLEO}, the $K^0_S \rho$ mode is chosen to have unit 
amplitude and zero phase.

For Belle, a set of 18 two-body amplitudes is used. These include five
Cabibbo-allowed amplitudes: $K^{*}(892)^{+}\pi^-$, $K^{*}(1410)^{+}\pi^-$,
$K^{*}_0(1430)^{+}\pi^-$, $K^{*}_2(1430)^{+}\pi^-$ and $K^*(1680)^+\pi^-$, 
their doubly Cabibbo-suppressed partners, and eight channels with a $K^0_S$ 
and a $\pi\pi$ resonance: $\rho$, $\omega$, $f_0(980)$, $f_2(1270)$, 
$f_0(1370)$, $\rho(1450)$, $\sigma_1$ and $\sigma_2$ . 
The  Breit--Wigner  masses and widths of the scalars
$\sigma_1$ and $\sigma_2$ are left unconstrained, while the parameters of
the other resonances are taken to be the same as in the CLEO 
analysis~\cite{DALITZ_CLEO}. 
The parameters of the $\sigma$ resonances obtained in the fit are as 
follows: $M_{\sigma_1} = 519 \pm 6$ MeV/$c^2$, $\Gamma_{\sigma_1} = 
454 \pm 12$ MeV/$c^2$, $M_{\sigma_2} = 1050 \pm 8$ MeV/$c^2$ and  
$\Gamma_{\sigma_2} = 101 \pm 7$ MeV/$c^2$ (the errors are statistical only).
In the BaBar case, a similar model is used with 16 two-body decay amplitudes
and phases. In particular, a model based on a fit to scattering data (K-matrix \cite{ref:AS}) is used to parametrize alternatively the $\pi \pi$ S-wave component and it is used to estimate the model systematic uncertainty. The agreement between the data and the fit result is satisfactory 
for the purpose of measuring $\gamma$ and the discrepancy is taken into 
account in the model uncertainty.

Once $f(m^2_+, m^2_-)$ is determined, a fit to $B^{\pm}$ data allows the determination 
of $r_B$, $\gamma$ and $\delta_B$, 
where  $\delta_B =\arg [A(B^+ \to \overline{D} K^+)/A(B^+ \to  D K^+)]$. 
Analysis of $CP$ violation is performed by means of an unbinned maximum 
likelihood fit with the $B^+$ and $B^-$ samples fitted separately using 
Cartesian parameters $x_{\pm} = r_B^{\pm} \cos (\delta_B\pm \gamma)$ and
$y_{\pm} = r_B^{\pm} \sin (\delta_B\pm \gamma )$. The fit is performed 
by minimizing the negative likelihood function of $n$ events
\begin{equation}
-2 \log L = -2 \sum_{i=1}^{n} \log p(m^2_{+,i}, m^2_{-,i}, \Delta E_i,
M_{bc, i}),
\end{equation}
with the Dalitz plot density $p$ represented as 
%
\begin{equation}
\begin{split}
p(&m_+^2,m_-^2,\Delta E,M_{\rm bc}) = \\
&\epsilon  | f(m_+^2,m_-^2) + (x +i y) f(m_-^2,m_+^2)|^2\times  \\
&\times F_{\rm sig}(\Delta E, M_{\rm bc}) 
+ F_{\rm bck}(m_+^2,m_-^2,\Delta E,M_{\rm bc}).
\end{split}
\end{equation}
The signal distribution $F_{\rm sig}$ is a function of two
kinematic variables, $\Delta E$ and $M_{\rm bc}$, $F_{\rm bck}$ is the 
distribution of the background, and $\epsilon = \epsilon(m_+^2,m_-^2)$
is the total efficiency.
The background density function $F_{\rm bck}$ is determined from
analysis of sideband events in data and with MC generated events.

\begin{figure}[htb]
\begin{center}
\includegraphics[width=15pc]{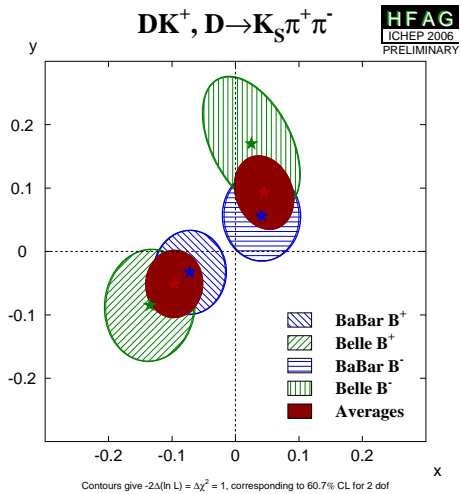}
\end{center}
\caption{Results of signal fits with free parameters $x_{\pm} = 
r \cos \pm \gamma + \delta_B $ and $y_{\pm} = r \sin \pm \gamma + \delta_B$ for 
$B^\pm \to D K^\pm$ from the BaBar and Belle latest publications~\cite{HFAG}. 
The contours indicate one standard deviation.}
\label{fig:xy_dalitz}
\end{figure}
Figure~\ref{fig:xy_dalitz} shows the results of the separate $B^+$ and
$B^-$ data fits for $B \to D K$ mode in the $x$--$y$ plane for the BaBar 
and Belle collaborations. 
Confidence intervals were then calculated using a frequentist technique
(the so-called Neyman procedure in the BaBar case, the unified approach
of Feldman and Cousins~\cite{FC} in the Belle case). The central values
for the parameters $\gamma$, $r_B$ and $\delta$ for the combined fit 
(using the $(x_{\pm}, y_{\pm})$ obtained for all modes) with their one-standard-deviation intervals are presented in Tab.~\ref{table:2} for the BaBar and Belle analysis.
  
 The uncertainties in the model used to parametrize the 
$\bar{D}^0 \to K^0_S \pi^+ \pi^-$ decay amplitude lead to an
associated systematic error in the fit result.
These uncertainties arise from the fact that there is no unique
choice for the set of quasi-2-body channels in the decay, as 
well as from the various possible parameterizations 
of certain components, such as the non-resonant amplitude. 
To evaluate this uncertainty several alternative models
have been used to fit the data. 
\begin{table*}[htb]
\caption{Results of the combination of $B^+ \to D K^+$, $B^+ \to D^* K^+$,
and $B^+ \to D K^{*+}$ modes for BaBar and Belle analyses. 
The first error is statistical, the
second is systematic and the third one is the model error.
In the case of BaBar, one standard deviation constraint is given for 
the $r_B$ values.}
\label{table:2}
\newcommand{\m}{\hphantom{$-$}}
\newcommand{\cc}[1]{\multicolumn{1}{c}{#1}}
\renewcommand{\tabcolsep}{2pc} 
\renewcommand{\arraystretch}{1.2} 
\begin{center}
\begin{tabular}{@{}lcc}
\hline
Parameter & BaBar & Belle \\
\hline
$\gamma$ & $(92 \pm 41 \pm 11 \pm 12)^\circ$ & 
           $(53^{+15}_{-18} \pm 3 \pm 9)^\circ$ \\
$r^{DK}_B$ & $< 0.140$ & 
$0.159^{+0.054}_{-0.050} \pm 0.012 \pm 0.049$ \\
$\delta^{DK}_B$ & 
$(118 \pm 63 \pm 19 \pm 36)^\circ$ & 
$(146^{+19}_{-20} \pm 3 \pm 23)^\circ$ \\
$r^{D^*K}_B$ & $0.017 - 0.203$ & 
$0.175^{+0.108}_{-0.099} \pm 0.013 \pm 0.049$ \\
$\delta^{D^*K}_B$  & 
$(-62 \pm 59 \pm 18 \pm 10)^\circ$ & 
$(302^{+34}_{-35} \pm 6 \pm 23)^\circ$ \\
$r^{DK^*}_B$ & & 
$0.564^{+0.216}_{-0.155} \pm 0.041 \pm 0.084$ \\
$\delta^{DK^*}_B$ & & 
$(243^{+20}_{-23} \pm 3 \pm 49)^\circ$ \\
\hline
\end{tabular}
\end{center}
\end{table*}

Despite similar statistical errors obtained for $(x_{\pm}, y_{\pm})$ in 
the two experiments, the resulting $\gamma$ error is much smaller in 
Belle's analysis.
Since the uncertainty on $\gamma$ scales roughly as $1/r_B$, the difference
is explained by noticing that the BaBar $(x_{\pm}, y_{\pm})$ measurements
favor values of $r_B$ smaller than the Belle results. 

At present the amplitude in the Dalitz plot analysis is described as a sum
of Breit--Wigner-like resonances \eqref{ffunc}. This approach
is valid for narrow well-spaced resonances but fails to
describe broad resonances, in particular the scalar ones. In addition,
interferences between overlapping resonances may not be well accounted
for within the Breit--Wigner model, which in turn can have an impact on the determination of the 
CKM parameter $\gamma$. The $K$-matrix approach appears as a possible alternative that correctly
implements unitarity of $S$ matrix in 2-body scattering also for overlapping resonances. Its extension to 3-body decays
is delicate with incomplete analytic structure from unitarity constraints. Nevertheless,  
a $K$-matrix approach extended to 3-body decays would provide an
alternative to the current model (sum of Breit--Wigner-like resonances)
and help to assess the model error more precisely \cite{descotes}.

The error due to the resonance model can be avoided by using the model-independent 
$\gamma$ measurement proposed in \cite{GGSZ}. In this approach, 
the Dalitz plot is partitioned in bins symmetric  with
respect to the $\pi^+ \pi^-$ axis. Counting the number of events in such bins from entangled $D$ decay samples, in addition to the already
utilized flavour-tagged $D$ decay samples,
 can determine the strong phase variation over the Dalitz plot.
 For this the data of a $\tau$-charm factory is needed. 
Useful samples consist of $\psi(3770) \to D^0 \overline{D}^0$ events 
where one of the $D$ mesons decays into a $CP$ eigenstate (such as $K^+ K^-$ 
or $K_S^0 \omega$), while the $D$ meson going in the opposite direction decays into $K_S^0 \pi^+ \pi^-$.  Using also a similar sample where both mesons from the $\psi(3770)$ decay into the $K^0  \pi^+ \pi^-$ state provides enough information to measure all the needed hadronic parameters in $D$ decay up to one overall discrete ambiguity (this can be resolved using a Breit-Wigner model). CLEO-c showed that  with the current 
integrated luminosity of 280 pb$^{-1}$ at the $\psi(3770)$ resonance,
these samples  are already available.

With the luminosity of 750 pb$^{-1}$, that CLEO-c should get at  the end of its operation,
the samples will be respectively about $1000$ and $2000$ events.
Using these two samples with a binned analysis and assuming $r_B = 0.1$,
a  $4^o$ precision on $\phi_3$  could be  obtained \cite{Bondar:2005ki,Bondar:2007ir}. An unbinned implementation of the model independent approach was
 presented by A. Poluektov \cite{Bondar:2007ir}.  
 
 Channels with bigger $r_B\sim 0.3-0.4$, such as $B^0 \to D^0 K^{*0}$, have been proposed. An  analysis of this channel exploits the $b$-quark flavour tag provided by the sign of the charged kaon in the  $ K^{*0}$ decay \cite{sordini}.

\section{Measurements of $\sin(2\beta+\gamma)$ }
  A $B^0$ meson can decay into $D^{-(*)}\pi^{+}$ final state either directly  through a 
  Cabibbo-favoured transition  (proportional to $V_{cb}$)  or can first oscillate into a $\bar B^0$ and then decay via    
  a doubly Cabibbo-suppressed transition (proportional to $V_{ub}$).  The interference of the two contributions generates
  the observables $S^{\pm}$ in the time-dependent CP asymmetries that are equal to  $2 r^{(*)}\sin(2\beta+\gamma\pm \delta)$ \cite{BDpi,RF-gam-ca}, 
  where $ r e^{i\delta}={A( \bar B^0 \rightarrow D^- \pi^+)}/{A(   B^0 \rightarrow D^- \pi^+)}$. Unfortunately this ratio is very small, $O(0.02)$, 
  and one furthermore needs to have knowledge of the relative strong phase $\delta$ in order to be able to extract the weak phases. To do so
  one either needs to measure the observables with $O(r^2)$ precision or use external input on $r$.

\begin{figure}[htb]
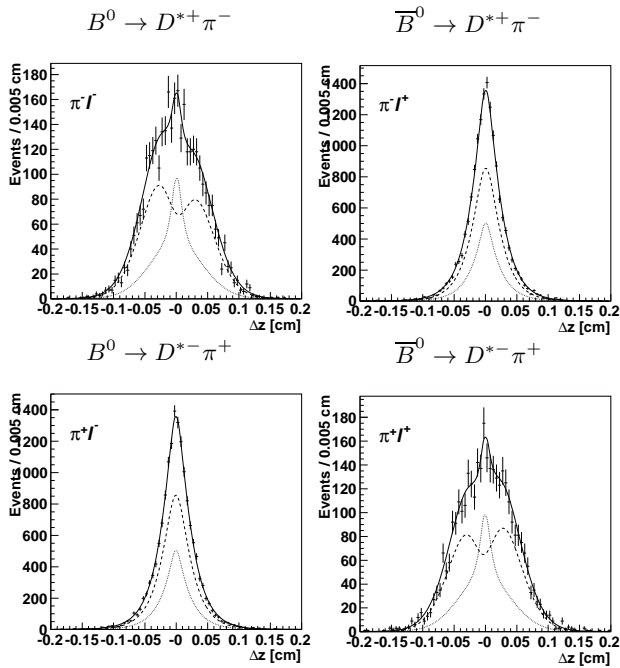

\begin{minipage}{4cm}
\begin{center}
$B^0 \to D^{*+}\pi^-$\\
\vspace*{0.4cm}
\resizebox{4cm}{!}{\includegraphics
      {fig3a.epsi}}
\end{center}
\end{minipage}
\begin{minipage}{4cm}
\begin{center}
$\overline{B}^0 \to D^{*+}\pi^-$\\
\vspace*{0.4cm}
\resizebox{4cm}{!}{\includegraphics
      {fig3b.epsi}}
\end{center}
\end{minipage}

\begin{minipage}{4cm}
\begin{center}
$B^0 \to D^{*-}\pi^+$\\
\vspace*{0.4cm}
\resizebox{4cm}{!}{\includegraphics
      {fig3c.epsi}}
\end{center}
\end{minipage}
\begin{minipage}{4cm}
\begin{center}
$\overline{B}^0 \to D^{*-}\pi^+$\\
\vspace*{0.4cm}
\resizebox{4cm}{!}{\includegraphics
      {fig3d.epsi}}
\end{center}
\end{minipage}
\caption{\label{fig:dt-part}
$\Delta t$ distributions for the partial reconstruction
 sample from Belle experiment. Curves are the fit results for the signal, background, and
 their sum.}
\end{figure}

   BaBar and Belle have performed time-dependent analyses with full and partial reconstruction techniques (for the $D^{-*}\pi^{+}$ channel, see Fig.~\ref{fig:dt-part})  \cite{ganzhur,abe}, giving 
     \begin{equation}
   a^{D^*\pi} = 2 r^* \sin{2\beta + \gamma} \cos{\delta} = -0.037 \pm 0.011 ,
    \label{eq:adstarpi}
  \end{equation}
with an error that is still dominated by the statistical component.

Estimate of $r$ in $B^0 \rightarrow D^- \pi^+$ from the $B^0 \rightarrow D_s^- \pi^+$ decay using $SU(3)$ symmetry was presented by M. Baak \cite{baak}.
The potential breaking of underlying assumptions can come from several sources: non-factorizable contributions, final state interactions, or missing diagrams in calculation - e.g. W-exchange. Nevertheless a global fit to several observables can constrain such effects, which leaves hopes that such measurement can be included in the Unitarity Triangle global fits ~\cite{baak}.

Interesting ideas on how to extract $\sin(2\beta+\gamma)$ from multi-body decays have been discussed. In particular, a time-dependent Dalitz analysis 
 of the $B^0 \to D^- K^0 \pi^+$ decay can separate $V_{cb}$ and $V_{ub}$ contributions (visible through $K^*$ and  $D^{**}$ resonances respectively)  and therefore be sensitive to the weak phase. Unfortunately, given the level of background  only with 10 ab$^{-1}$ of integrated luminosity one can aim at a 10\% error \cite{polci}.

\section{$\beta$ and $\gamma$ at hadron  colliders.}

An overview of various $\gamma$ determinations using $B_s$ decays into charmed final states  was given by R. Fleischer \cite{Fleischer:2007wc}. For the decays that have both tree and penguin amplitudes, the $U$-spin symmetry is used to obtain the information on the penguin-to-tree ratio. In the $U$-spin based methods only the $SU(3)$ flavour 
symmetry is used, while in other uses of $SU(3)$, for instance in diagrammatic approaches, further dynamical assumptions such as neglecting annihilation-like amplitudes are commonly used. The $U$-spin symmetry offers also a powerful
tool for the analysis of the $B_d\to D^{(*)\pm} \pi^\mp$, $B_s\to D_s^{(*)\pm} K^\mp$
system \cite{RF-gam-ca}.

The hadronic matrix elements of the $B_{s}\to J/\psi K_{\rm S}$ and 
$B_{d}\to J/\psi K_{\rm S}$ decays are related through the $U$-spin symmetry 
\cite{RF-BspsiK}. The penguin and tree amplitudes in $B_{s}\to J/\psi K_{\rm S}$ 
are multiplied by the combinations of CKM elements of similar size, $V_{cb}^*V_{cd}$ and $V_{ub}^*V_{ud}$ respectively. In $B_{d}\to J/\psi K_{\rm S}$, on the other hand, the 
tree is relatively $\sim 1/\lambda^2$ enhanced compared to the penguin. This hierarchy allows for the determination of penguin pollution on $\sin 2 \beta$ determination for both decays simultaneously, up to the $SU(3)$-breaking effects, thereby complementing
the discussion given in Section~\ref{sec:beta}. This type of analysis can also be used 
to determine the hadronic penguin effects in the extraction of the 
$B^0_s$--$\bar B^0_s$ mixing phase $\phi_s\stackrel{\rm SM}{\approx} -2^\circ$ 
from the $B_s\to J/\psi \phi$ channel \cite{DDFN} by relating it to the 
$B_d\to J/\psi \rho^0$ decay \cite{RF-ang}. 

Another interesting $U$-spin-related  
system is given by the $B_s\to D_s^+D_s^-$ and $B_d\to D^+D^-$ 
decays \cite{RF-BspsiK,RF-BDD}. Here we may take the penguin effects into
account in the determination of the $B^0_d$--$\bar B^0_d$ and
$B^0_s$--$\bar B^0_s$ mixing phases $\phi_d$ and $\phi_s$, respectively. 
As was noted in Ref.~\cite{habil}, the analysis of the 
$B_{d(s)}\to D^+_{d(s)}D^-_{d(s)}$ decays can also 
straightforwardly be applied to the $B_{d(s)}\to K^0 \bar K^0$ system. Following 
these lines, the penguin effects in the determination of
$\sin\phi_s$ from the $b\to s$ penguin decay $B^0_s\to K^0\bar K^0$
can be included through its $B^0_d\to K^0\bar K^0$ partner \cite{CPS-NP} (here
$B^0_s\to K^0\bar K^0$ and $B^0_d\to K^0\bar K^0$ take the r\^oles of 
$B^0_s\to D^+_sD^-_s$ and $B^0_d\to D^+D^-$, respectively); this is also 
the case for the corresponding $B_{d(s)}\to K^{*0} \bar K^{*0}$ decays 
\cite{RF-ang, matias}. 

The theoretically cleanest determinations of the mixing phases $\phi_s$ and $\beta$ are offered by the pure tree decays $B_s\to D_\pm K_{\rm S(L)}$ and $B_d\to D_\pm\pi^0, D_\pm\rho^0, ...$, respectively \cite{RF-BCP}. The weak phase $\gamma$, on the other hand, can be obtained from pure tree colour-allowed 
$\Delta S=1$ decays $B_s\to D_s^\pm K^\mp$ \cite{BsDsK}
and/or from the pure tree colour-suppressed $\Delta S=1$  $B_s\to D\eta^{(')}$, 
$B_s\to D\phi$, ...\ and $B_d\to D K_{\rm S(L)}$ decays. Since these are tree decays there is no penguin pollution. There is enough experimental information to extract all the hadronic parameters because many different $D$ decays can be used for the same $B$ decay process. Each additional $D$ decay mode brings in one additional parameter, the strong phase between $D$ and $\bar D$ decay, while also bringing in two additional observables, the corresponding branching ratio and the CP asymmetry.

The  study of $B^{-} \to D^{0}K^{-}$ decay  by CDF, where the $D^{0}$ is reconstructed in flavor ($K^{-}\pi^{+}$) or CP-even ($K^{-}K^{+}, \pi^{-}\pi^{+}$) eigenstates was reported \cite{rahaman}, with the measurement of the ratio  $R = {BR(B^{-} \rightarrow D^{0}_{flav}K^{-})}/{BR(B^{-} \rightarrow D^{0}_{flav}\pi^{-})}$, which is one of the inputs in the GLW method for $\gamma$ determination \cite{GLW}, 
quoting the value $0.065\pm0.007\pm0.004$ \cite{bib:r2}. 

CDF  observed for the first time the $B_{s}^{0}~\rightarrow D_{s}^{+}D_{s}^{-}$ 
channel and reported the measurement of the ratio 
$R = {BR(B_{s} \rightarrow D_{s}^{+}D_{s}^{-})}/{BR(B^{0} \rightarrow 
D_{s}^{+}D^{-})}$=$1.67 \pm 0.41\mbox{(stat)} \pm 0.12\mbox{(syst)} 
\pm 0.24(f_{s}/f_{d}) \pm 0.39(Br_{\phi\pi})$ \cite{CDF-BsDsDs}. Performing a 
run on the $\Upsilon(5S)$ resonance, also the Belle collaboration has
recently obtained an upper bound of $6.7\%$ (90\% C.L.) for this branching ratio
\cite{Belle-Y5S}. Moreover, the D0 collaboration has performed a first
analysis of the combined $B_s\to D_s^{(*)}D_s^{(*)}$ branching ratio, with 
the result of $\mbox{BR}(B_s\to D_s^{(*)}D_s^{(*)})=(3.9^{+1.9+1.6}_{-1.7-1.5})\%$
\cite{D0-BsDsDs}. For a recent analysis using these Tevatron results to control
the penguin effects in $B^0_d\to D^+D^-$ see Ref.~\cite{RF-BDD}.

The LHCb sensitivity for the extraction of $\gamma$ was simulated for 
$B^\pm\to DK^\pm$ tree-level decays \cite{patel}. A combination of the GLW and ADS methods with the flavour $D\to K\pi$ and $D\to K\pi\pi\pi$ decays leads to a statistical error on $\gamma$ in the range $5$--$12^\circ$ for $r_B\sim 0.08$ with $2$ fb${}^{-1}$ data. The use of $B^\pm \to D^{*0}K^\pm$ that are more challenging at LHCb is also under study. The statistical precision on $\gamma$ from neutral $B$ decays has been estimated to be in the range $7$--$10^\circ$ for $2$ fb${}^{-1}$ of LHCb data, while a Dalitz analysis in $B^\pm\to (K_S\pi^+\pi^-)_DK^\pm$ is estimated to lead to a statistical error on $\gamma$ of about $8^\circ$. An impact of the four-body $D$ decay, $B^\pm\to (K^+K^-\pi^+\pi^-)_DK^\pm$ was also simulated, with estimated accuracy $\gamma\sim 14^\circ$. All in all the estimated precision on $\gamma$ from a combination of these modes is expected to be at $\sim 5^\circ$ for $2$ fb${}^{-1}$, which is comparable to the indirect determination of $\gamma$ using CKM fits.

\section{Conclusions}
In the next two years, the $e^+e^-$  $B$-factories will reach a total integrated 
luminosity of about 2 ab$^{-1}$ and CDF/D0 of several  fb$^{-1}$. The measurement 
of the angle $\beta$ will be performed in several channels with no limitation due to systematics uncertainty and with a theory error under control.  The current world 
average error on $\gamma$ is around $20^\circ$ \cite{marti,trabelsi}. 
  A  more precise  measurement  will be challenging, especially since the 
sensitivity depends critically on the real value of $r_B$ for the various channels  
that need to be combined. Thanks to the quickly approaching start of the LHC and
its dedicated $B$-decay experiment LHCb, we will soon get full access to the
rich physics potential of the $B_s$-meson system, and will also enter a new era 
for the precision measurements of $\gamma$. In the more distant future, an 
upgrade of LHCb and a super $B$-factory (or a super flavour factory) could 
bring the measurements to their ultimate precisions.


\begin{acknowledgments}
 The authors are most grateful for the excellent organization of the CKM 
 Workshop in Nagoya, allowing us to run the sessions very smoothly,
 which helped us to have very productive discussions. The work of J.Z. is supported in 
part by the European Commission RTN network, Contract No.~MRTN-CT-2006-035482 
(FLAVIAnet) and by the Slovenian Research Agency. 
\end{acknowledgments}


\begin{thebibliography}{99}


\bibitem{bib:C}
  N.~Cabibbo, Phys. Rev. Lett. {\bf 10}, 531 (1963).

\bibitem{bib:KM}
  M.~Kobayashi and T.~Maskawa, Prog. Theor. Phys. {\bf 49}, 652 (1973).


\bibitem{CKM05}G.~Cavoto {\it et al.},                                                                                         
 arXiv:hep-ph/0603019.                                                                                         

\bibitem{bisa}A.~B.~Carter and A.~I.~Sanda,
{ Phys.\ Rev.\ Lett.}~{\bf 45}, 952 (1980);
{ Phys.\ Rev.}\ D {\bf 23}, 1567 (1981);

I.~I.~Bigi and A.~I.~Sanda,
{ Nucl.\ Phys.}\ B {\bf 193}, 85 (1981).

\bibitem{RF-rev}For a detailed discussion, see 
R.~Fleischer,
  J.\ Phys.\ G {\bf 32} (2006) R71
  [arXiv:hep-ph/0512253].

\bibitem{BMR} H.~Boos, T.~Mannel and J.~Reuter,
  Phys.\ Rev.\  D {\bf 70} (2004) 036006
  [arXiv:hep-ph/0403085].

\bibitem{LM}H.-n.~Li and S.~Mishima,
  JHEP {\bf 0703} (2007) 009
  [arXiv:hep-ph/0610120].
  
\bibitem{ref:mishima}
  S.~Mishima,
  arXiv:hep-ph/0703193.

  
\bibitem{CPS}M.~Ciuchini, M.~Pierini and L.~Silvestrini,
  Phys.\ Rev.\ Lett.\  {\bf 95} (2005) 221804
  [arXiv:hep-ph/0507290].
  
\bibitem{ref:ciuchini}
 M.Ciuchini, proc. this workshop.
  
\bibitem{RF-BspsiK}R.~Fleischer,
  Eur.\ Phys.\ J.\  C {\bf 10} (1999) 299
  [arXiv:hep-ph/9903455].

\bibitem{BF-06}P.~Ball and R.~Fleischer,
  Eur.\ Phys.\ J.\  C {\bf 48} (2006) 413
  [arXiv:hep-ph/0604249].
  
  \bibitem{ball}  P.~Ball,
  arXiv:hep-ph/0612325.
  
  
\bibitem{NP-Bd}M.~Bona {\it et al.}  [UTfit Collaboration],
  JHEP {\bf 0603}, 080 (2006)
  [arXiv:hep-ph/0509219];
A.~J.~Buras, R.~Fleischer, S.~Recksiegel and F.~Schwab,
  Eur.\ Phys.\ J.\  C {\bf 45} (2006) 701
  [arXiv:hep-ph/0512032];
M.~Blanke, A.~J.~Buras, D.~Guadagnoli and C.~Tarantino,
  JHEP {\bf 0610}, 003 (2006)
  [arXiv:hep-ph/0604057];
M.~Bona {\it et al.}, The UT{\it fit} Collaboration, hep-ph/0605213.

\bibitem{kathy}
See for instance,
  K.~A.~George,
  arXiv:hep-ex/0701044.

\bibitem{bellesin2beta}
  K.-F.~Chen {\it et al.}  [Belle Collaboration],
  Phys.\ Rev.\ Lett.\  {\bf 98}, 031802 (2007)
  [arXiv:hep-ex/0608039].
 
\bibitem{Datta:2007mj}
  A.~Datta,
  arXiv:hep-ph/0702270.
 
\bibitem{DDF-ambig}A.~S.~Dighe, I.~Dunietz and R.~Fleischer,
  Phys.\ Lett.\  B {\bf 433} (1998) 147
  [arXiv:hep-ph/9804254].

 \bibitem{prd1}
 B.~Aubert {\it et al.}  [BABAR Collaboration],
  Phys.\ Rev.\  D {\bf 71}, 032005 (2005)
  [arXiv:hep-ex/0411016].

  

\bibitem{browder}
  T.~E.~Browder, A.~Datta, P.~J.~O'Donnell and S.~Pakvasa,
  Phys.\ Rev.\  D {\bf 61}, 054009 (2000)
  [arXiv:hep-ph/9905425];
For earlier papers see:J.~Charles, A.~Le Yaouanc, L.~Oliver, O.~Pene and 
J.~C.~Raynal,
  Phys.\ Lett.\  B {\bf 425}, 375 (1998)
  [Erratum-ibid.\  B {\bf 433}, 441 (1998)]
  [arXiv:hep-ph/9801363];
P.~Colangelo, F.~De Fazio, G.~Nardulli, N.~Paver and Riazuddin,
  Phys.\ Rev.\  D {\bf 60}, 033002 (1999)
[arXiv:hep-ph/9901264].

  

\bibitem{babarDDK}
  B.~Aubert {\it et al.}  [BABAR Collaboration],
  Phys.\ Rev.\  D {\bf 74}, 091101 (2006)
  [arXiv:hep-ex/0608016].
  
\bibitem{bondar}
  A.~Bondar, T.~Gershon and P.~Krokovny,
  Phys.\ Lett.\  B {\bf 624}, 1 (2005)
  [arXiv:hep-ph/0503174].

 \bibitem{gaz}
 A.~Gaz,  proc. this workshop.
 
\bibitem{RF-BDD}R.~Fleischer,
arXiv:0705.4421 [hep-ph], to appear in Eur.\ Phys.\ J {\bf C}
  
\bibitem{justin} 
  J.~Albert, A.~Datta and D.~London,
  Phys.\ Lett.\  B {\bf 605}, 335 (2005)
  [arXiv:hep-ph/0410015];
A.~Datta and D.~London,
  Phys.\ Lett.\  B {\bf 584}, 81 (2004)
  [arXiv:hep-ph/0310252].
  
 \bibitem{fratina}
 S.~Fratina,  proc. this workshop.

\bibitem{Chen:2005rp}
  C.~H.~Chen, C.~Q.~Geng and Z.~T.~Wei,
  Eur.\ Phys.\ J.\  C {\bf 46}, 367 (2006)
  [arXiv:hep-ph/0507295].

\bibitem{Covarelli:2007cn}
  R.~Covarelli,
  arXiv:hep-ex/0702040.  

  

\bibitem{ciuch}
  M.~Ciuchini {\it et al.}
  JHEP {\bf 0308}, 031 (2003)
  [arXiv:hep-ph/0308029].

\bibitem{bene}
M.~Beneke {\it et al.}, Phys.\ Lett. {\bf B 576}, 173 (2003)  [arXiv:hep-ph/0307344].



\bibitem{fmv}
  B.~Aubert {\it et al.}  [BABAR Collaboration],
  Phys.\ Rev.\  D {\bf 70}, 012007 (2004)
  [arXiv:hep-ex/0403002].


\bibitem{bozzi}
  B.~Aubert {\it et al.}  [BABAR Collaboration],
  Phys.\ Rev.\ Lett.\  {\bf 88}, 231801 (2002)
  [arXiv:hep-ex/0202041].





\bibitem{koste}
  V.~A.~Kostelecky,
  Phys.\ Rev.\ Lett.\  {\bf 80}, 1818 (1998)
  [arXiv:hep-ph/9809572].



 




%








\bibitem{GLW}  B. Carter and A. I. Sanda, Phys. Rev. D 23, 1567 (1981); M.~Gronau and D.~London, Phys.\ Lett.\ B {\bf 253},  483 (1991);
M.~Gronau and D.~Wyler, Phys.\ Lett.\ B {\bf 265},  172 (1991);
M.~Gronau, Phys.\ Rev.\ D {\bf 58}, 037301  (1998) [arXiv:hep-ph/9802315].

\bibitem{ADS} D.~Atwood, I.~Dunietz and A.~Soni, Phys.\ Rev.\ Lett.\ {\bf 78}, \
3257 (1997) 
  [arXiv:hep-ph/9612433].



\bibitem{GGSZ}  
  A.~Giri, Y.~Grossman, A.~Soffer and J.~Zupan,
  Phys.\ Rev.\  D {\bf 68}, 054018 (2003)
  [arXiv:hep-ph/0303187].

\bibitem{GRONAU} 
  M.~Gronau,
  Phys.\ Lett.\  B {\bf 557}, 198 (2003)
  [arXiv:hep-ph/0211282].

\bibitem{Gronau:2004gt}
  M.~Gronau, Y.~Grossman, N.~Shuhmaher, A.~Soffer and J.~Zupan,
  Phys.\ Rev.\  D {\bf 69}, 113003 (2004)
  [arXiv:hep-ph/0402055];
  D.~Atwood and A.~Soni,
  Phys.\ Rev.\  D {\bf 71}, 013007 (2005)
 [arXiv:hep-ph/0312100].

\bibitem{Grossman:2005rp}
  Y.~Grossman, A.~Soffer and J.~Zupan,
  Phys.\ Rev.\  D {\bf 72}, 031501 (2005), 
  [arXiv:hep-ph/0505270]; see also
  M.~Gronau, Y.~Grossman, Z.~Surujon and J.~Zupan,
  Phys.\ Lett.\  B {\bf 649}, 61 (2007),
  [arXiv:hep-ph/0702011].


\bibitem{DALITZ_BELLE} 
  A.~Poluektov {\it et al.}  [Belle Collaboration],
  Phys.\ Rev.\  D {\bf 73}, 112009 (2006)
  [arXiv:hep-ex/0604054].



\bibitem{DALITZ_BABAR_NEW} 
  B.~Aubert {\it et al.}  [BABAR Collaboration],
  arXiv:hep-ex/0607104.



\bibitem{DALITZ_BABAR_OLD} 
  B.~Aubert {\it et al.}  [BABAR Collaboration],
  Phys.\ Rev.\ Lett.\  {\bf 95}, 121802 (2005)
  [arXiv:hep-ex/0504039].



\bibitem{DALITZ_CLEO} 
  H.~Muramatsu {\it et al.}  [CLEO Collaboration],
  Phys.\ Rev.\ Lett.\  {\bf 89}, 251802 (2002)
  [Erratum-ibid.\  {\bf 90}, 059901 (2003)]
  [arXiv:hep-ex/0207067].
\bibitem{ref:AS}  V.V. Anisovich and A.V. Sarantsev, Eur.\ Phys.\ Jour. {\bf A16}, 229 (2003).



\bibitem{HFAG} Heavy Flavor Averaging Group:
www.slac.stanford.edu/xorg/hfag/.
\bibitem{FC} 
  G.~J.~Feldman and R.~D.~Cousins,
  Phys.\ Rev.\  D {\bf 57}, 3873 (1998)
  [arXiv:physics/9711021].


\bibitem{sordini} V.Sordini, 
proc this workshop, hep-ph/0703292

\bibitem{descotes} S. Descotes-Genon, proc. this workshop.

\bibitem{Bondar:2005ki}
  A.~Bondar and A.~Poluektov,
  Eur.\ Phys.\ J.\  C {\bf 47}, 347 (2006)
  [arXiv:hep-ph/0510246].
  
\bibitem{Bondar:2007ir}
  A.~Bondar and A.~Poluektov,
  arXiv:hep-ph/0703267.


\bibitem{BDpi}I.~Dunietz and R.~G.~Sachs,
{ Phys.\ Rev.}\ D {\bf 37}, 3186 (1988) [Erratum-ibid.\ D {\bf 39}, 3515 (1989)];
I.~Dunietz,
{ Phys.\ Lett.}\ B {\bf 427}, 179 (1998) [arXiv:hep-ph/9712401];
D.~A.~Suprun, C.~W.~Chiang and J.~L.~Rosner,
{ Phys.\ Rev.}\ D {\bf 65}, 054025 (2002) [arXiv:hep-ph/0110159].

\bibitem{RF-gam-ca}R. Fleischer,
 {  Nucl.\ Phys.}\ B {\bf 671}, 459 (2003)
  [arXiv:hep-ph/0304027].


 \bibitem{abe}
 K.~Abe,  proc. this workshop


 \bibitem{ganzhur}
 S.~Ganzhur,  proc. this workshop
 

 \bibitem{baak}
 M.~Baak,  proc. this workshop
 
 
 \bibitem{polci}
 F.~Polci,  proc. this workshop, hep-ex/0703047

\bibitem{Fleischer:2007wc}
  R.~Fleischer,
  arXiv:hep-ph/0701216.

\bibitem{DDFN}
A.~S.~Dighe, I.~Dunietz and R.~Fleischer,
{ Eur.\ Phys.\ J.}\ C {\bf 6}, 647 (1999) [arXiv:hep-ph/9804253];
I. Dunietz, R. Fleischer and U. Nierste,
  { Phys.\ Rev.}\ D {\bf 63}, 114015 (2001)
  [arXiv:hep-ph/0012219].


\bibitem{RF-ang} R.~Fleischer,
  Phys.\ Rev.\ D {\bf 60}, 073008 (1999)
  [arXiv:hep-ph/9903540].

\bibitem{habil}R.~Fleischer,
  {\it Phys.\ Rept.}~{\bf 370} (2002) 537 [arXiv:hep-ph/0207108].

\bibitem{CPS-NP}M.~Ciuchini, M.~Pierini and L.~Silvestrini,
  hep-ph/0703137;
  
\bibitem{matias}S.~Descotes-Genon, J.~Matias and J.~Virto,
  arXiv:0705.0477 [hep-ph].

\bibitem{RF-BCP} R.~Fleischer,
  Phys.\ Lett.\ B {\bf 562}, 234 (2003)
  [arXiv:hep-ph/0301255];
  Nucl.\ Phys.\ B {\bf 659}, 321 (2003)
  [arXiv:hep-ph/0301256].
  
\bibitem{BsDsK}R. Aleksan, I. Dunietz and B. Kayser,
{ Z.\ Phys.}\ C {\bf 54}, 653 (1992).

\bibitem{rahaman} A.~Rahaman, proc. this workshop. 

\bibitem{bib:r2}
  CDF collaboration, CDF public Note 7925 (2006).
  

  \bibitem{CDF-BsDsDs}S.~Farrington {\it et al.}\ [CDF Collaboration],
FERMILAB-CONF-06-405-E.

\bibitem{Belle-Y5S}K.~Abe {\it et al.}\  [Belle Collaboration],
  arXiv:hep-ex/0610003.
  
\bibitem{D0-BsDsDs}V.M.~Abazov {\it et a.}\ [D0 Coll.],
FERMILAB-PUB-07/047-E [hep-ex/0702049]




\bibitem{patel} M.~Patel, proc. this workshop 

\bibitem{marti} G.~Martinelli, proc. this workshop 
\bibitem{trabelsi} K.~Trabelsi, proc. this workshop 


\end{thebibliography}
\end{document}